\documentclass[prx,twocolumn,aps,showpacs,footinbib,10pt,numerical,superscriptaddress]{revtex4-1}
\usepackage[colorlinks=true,linkcolor=blue,urlcolor=blue,citecolor=blue]{hyperref}
\usepackage[english]{babel}
\usepackage[utf8]{inputenc}
\usepackage{amsmath}
\usepackage{amsfonts}
\usepackage{amssymb}
\usepackage{graphicx}
\usepackage{lmodern}
\usepackage{color}
\usepackage{enumerate}
\usepackage{mathtools}
\usepackage{hyperref}
\usepackage[normalem]{ulem}
\usepackage{bbm}
\usepackage{bbold}
\usepackage{float}
\usepackage[caption=false]{subfig}
\usepackage{enumitem}
\usepackage{verbatim}
\usepackage[]{placeins}
\usepackage{lipsum}
\usepackage{braket}
\usepackage{amsmath, bm}
\usepackage{kantlipsum}
\usepackage{graphicx,color}

\bibpunct{[}{]}{,}{n}{}{}

\begin{document}

\title{Majorana bound states in topological insulators with hidden Dirac points}
\author{Ferdinand~Schulz}
\author{Kirill Plekhanov}
\author{Daniel Loss}
\author{Jelena~Klinovaja}
\affiliation{
Department of Physics, University of Basel, Klingelbergstrasse 82, CH-4056 Basel, Switzerland}
\date{\today}

\begin{abstract}
{We address the issue whether it is possible to generate Majorana bound states at the magnetic-superconducting interface in two-dimensional topological insulators with hidden Dirac points in the spectrum. In this case, the Dirac point of edge states is located at the energies of the bulk states such that two types of states are strongly hybridized. Here, we show that well-defined Majorana bound states can be obtained even in materials with hidden Dirac point provided that the width of the magnetic strip is chosen to be comparable with the localization length of the edge states. The obtained topological phase diagram allows one to extract precisely the position of the Dirac point in the spectrum. In addition to standard zero-bias peak features caused by Majorana bound states in transport experiments, we propose to supplement future experiments with measurements of charge and spin polarization. In particular, we demonstrate that both observables flip their signs at the topological phase transition, thus, providing an independent signature of the presence of topological superconductivity. All features remain stable against substantially strong disorder.}\\
\vspace{1cm}
\end{abstract}

\maketitle

% ---------------------------------------------------------------

\section{Introduction}\label{seq.:sectionI}

The most prominent feature of two dimensional (2D) topological insulators (TIs) is the co-existence of perfectly conducting helical edge states at the boundary of the system with a gapped insulating bulk. Time-reversal symmetry (TRS) plays a crucial role to protect such a helical pair of gapless counter-propagating edge states~\cite{kane2005a, kane2005b, hassankane2010,  qizhang2011}. The first promising 2D TI material candidate were HgTe/CdTe and  InAs/GaSb  quantum wells~\cite{bhz2006,koenig2007, molenkamp2009,wangzhang2008,sullivan2010, sullivan2011,muraki_2013,klaus_2015,klaus_2017}. Despite the remarkable theoretical and experimental progress, handling of TIs remains a complicated task, with one of the main difficulties coming from the fact that the thickness of the quantum wells strongly affects the bandstructure of the system, and, moreover, sample inhomogeneities can result in trivial edge states  \cite{tr_1,tr_2}, which complicates the unambiguous detection of the topological phase.  Due to various reasons the conductance is never perfectly quantized even in the topological regime \cite{cond1,cond2,cond3,cond4,cond5,cond6,cond7,cond8,cond9,cond10,cond11,cond12}. Also, Josephson junction measurements are not always conclusive \cite{tr_3,tr_4,tr_5,tr_6}.  
Another unexpected puzzle
 is the fact that experimental studies of both quantum well systems show that the conductance values do not dependent 
 very noticeably on an externally applied magnetic field even if it is strong \cite{molenkamp2015, sullivan2015}. Such a behaviour is highly surprising, since the main effect of the magnetic field consists in breaking the TRS, leading theoretically to an opening of the gap at the Dirac point (DP) of the edge states of the TI. Hence, if the conductance is measured in the vicinity of the DP, which is supposedly well separated from the bulk states, it 
 should be strongly suppressed, in contrast to observations \cite{molenkamp2015, sullivan2015}. Recently, it was suggested that this surprising stability of the conductance quantization is related to a particular form of the bandstructure of the quantum wells, in which the DP is
hidden inside the bulk spectrum~\cite{songbo2014, wimmer2018,shen2018,ando}. 
This is an interesting assumption which we also adopt here and wish to explore in more detail.

\begin{figure}[!b]
\centering
\includegraphics[width=0.8\linewidth]{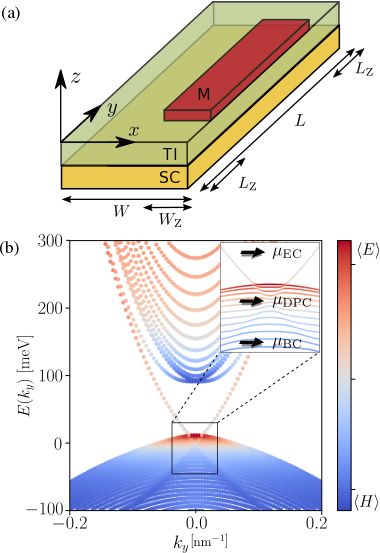}
\caption{(a) The setup consisting of a 2D TI (green slab) proximity coupled to an $s$-wave SC (yellow slab) with a magnetic strip (red slab) placed on one of edges. (b) The energy spectrum obtained numerically in a slab geometry of the TI with PBC along the $y$- and OBC along the $x$-direction. The color code refers to contributions of the $E$ and $H$ orbitals. Black arrows in the zoomed inset indicate specific positions of the chemical potential. Non-zero parameters are chosen to be $A=365$ meV/nm, $B=-686$ meV/nm$^2$, $D=-512$ meV/nm$^2$, $M=-40$ meV, $\mu_{\textrm{Edge}}=-100$ meV, $L_\text{Edge}/a=10$, $\mu=50$ meV, and $W/a=100$, with $a=1$ nm.}
\label{fig.:fullsetup}
\end{figure}

The hidden DP helps to stabilize the helical edge states even in the presence of external magnetic fields, which could be useful for many applications. However, such hidden DPs were argued to make it impossible to reach the topological phases hosting Majorana bound states (MBSs). MBSs are supposed to arise at the interface between superconductivity and magnetic dominating regions, if the chemical potential is tuned close to the DP. With the chemical potential being at the energy of the bulk states, this then brings also the continuum of bulk states into the play such that the standard scenario, based on energetically well-separated edge states, breaks down. Here, however, we show that this reasoning does not apply and that indeed
the presence of  a hidden DP does not hinder the appearance of MBSs.

We focus on the bandstructures of 2D TIs described by the Bernevig-Hughes-Zhang (BHZ) model~\cite{bhz2006}, where the DP is hidden inside the bulk  as a result of intrinsic bulk properties of the system (\textit{e.g.}, for particular values of the thickness of the quantum wells) or when the boundary of the sample has a generic edge potential~\cite{songbo2014,  wimmer2018}. By considering the TI being proximity-coupled to an $s$-wave superconductor (SC) as well  as in a contact with a magnetic strip inducing an effective Zeeman field on one of the edges [see Fig.~\ref{fig.:fullsetup}(a)], we propose to locate the position of the DP by using Majorana bound states (MBSs) as a detector. Even if the DP is hidden, the effective Zeeman field  needed to achieve the topological phase becomes minimal, if the chemical potential is placed at the DP. The resulting topological phase diagram is intrinsically related to the exact position of the DP, thus allowing one to find its energetic location. In our setup, a pair of MBSs is supported by the spatial interface between the superconducting and magnetic regions, at the two sides of the magnetic strip ~\cite{fukane2008,fukane2009,tanaka,sato,oppen,bjorn,jk2,keidel2018}. 
The standard detection method of MBSs is via transport measurements searching for zero-bias peaks in the conductance. However, such zero-bias peaks can also occur for other reasons such as Andreev bound states \cite{Pradareview}. To provide alternative detection methods for MBSs 
we propose to measure the local spin- and charge-polarization under the magnetic strip. As we will show, these two quantities undergo a sign flip at the topological phase transition, providing one more signature of the topological superconducting phase. In contrast to MBSs, which are observed locally, the spin and charge polarizations  originate from the bulk states and can be observed away from the interface. Importantly, the proposed effects are stable against weak disorder.

The present work is organized as follows. In Sec.~\ref{sec.:sectionII}, we introduce the model and explain the definition of the hidden DP in the spectrum of the 2D TI. In Sec.~\ref{sec.:sectionIII}, we present the topological phase diagram and compare it to the one obtained for an effective low-energy Hamiltonian in which bulk state contributions are neglect. In addition, we study MBS wavefunctions as a function of the chemical potential of the TI. In Sec.~\ref{sec.:sectionIV}, we focus on local bulk  properties that are changed across the topological phase transition. In particular, we show that the phase transition can be observed locally and away from the ends of the magnetic strip by measuring the spin or charge polarization. In Sec.~\ref{sec.:sectionV}, the stability of the system with respect to disorder is addressed as well as the role of the width of the magnetic strip.  Finally, we summarize our results and suggest possible experimental realizations. Additional details of numerical calculations are summarized in three appendices.

\section{Model}\label{sec.:sectionII}

We consider a 2D TI of a width $W$ (length $L$) along the $x$ ($y$) axis
[green slab in Fig.~\ref{fig.:fullsetup}(a)]. The description of the
TI is based on the BHZ model \cite{bhz2006}, defined on a square
lattice with lattice constant $a$. In momentum space
${\bf k}\equiv(k_x,k_y)$ and in the basis
$(c_{E\uparrow}, c_{H\uparrow}, c_{E\downarrow}, c_{H\downarrow})$,
the corresponding Hamiltonian can be written as
\begin{equation}
H_{\bf k}=\epsilon({\bf k})+M({\bf k})\tau_3+A[\sin (k_xa)\tau_1\sigma_3-\sin (k_ya)\tau_2],
\label{eqn:fullBHZ}
\end{equation}
where $\epsilon({\bf k})=-2D[2-\cos(k_xa)-\cos(k_ya)]$ and
$M({\bf k})=M-2B[2-\cos(k_xa)-\cos(k_ya)]$. Here, the Pauli matrices
$\tau_i$ and $\sigma_i$ act on orbital ($E,H$) and spin
($\uparrow,\downarrow$) degrees of freedom, 
%correspondingly, 
with
$c_{E/H\uparrow/\downarrow}$  being the corresponding annihilation operator.
The spin quantization axis is chosen to be along the $z$-direction.
In general, the parameters $A$, $B$, $D$, and $M$ depend on the
properties of the TI, determined by the material choice and by the
geometry of quantum wells. The parameters $B$ and $D$ are responsible
for a symmetric and an antisymmetric component of the effective masses
associated with different orbital degrees of freedom, while the
parameter $A$ determines the Fermi velocity. The parameter $M$,
responsible for the topological phase transition, flips its sign at a
critical quantum well thickness such that gapless helical states emerge in
a geometry with periodic boundary conditions (PBC) along the $x$-direction and with open boundary conditions (OBC) along the $y$-direction [see Fig.~\ref{fig.:fullsetup}(b)].

Previously, it was found that the intrinsic band structure of the
$8\times 8$ Kane Hamiltonian for a quantum well shows a natural emergence of the phenomenon that is referred to as
the `hidden DP'~\cite{songbo2014, wimmer2018}. The DP at zero momentum in the spectrum of edge states coincides in energy with bulk states, which is in contrast to more common models in which the DP is placed at the middle of the topological bulk gap.
In order to reproduce the same effect in the BHZ model, we tune the spectrum of the
helical edge states separately from the bulk spectrum by considering
the effect of a position-dependent chemical potential. More precisely,
we assume that the chemical potential in the bulk of the sample is
uniform and is given by $\mu$, while at the edge it is assumed to be
given by $\mu+\mu_\text{Edge}$ \cite{footnote}. This additional term $\mu_\text{Edge}$ allows us to reach the hidden DP regime, see
Fig.~\ref{fig.:fullsetup}(b). There, we denote by $L_{\textrm{Edge}}$ the
lengthscale associated with $\mu_\text{Edge}$, which is chosen to be
roughly the same as the decay length $\chi$ of the TI edge states, calculated at the DP in the absence of the edge potential (see
Appendix~\ref{Sec.:App1} for more details). We observe that, as a
result of such a symmetry breaking term, the DP moves away from the middle of
the topological band gap towards the energies of the conduction band, resulting in a 
hybridization with bulk states [see the inset in Fig.~\ref{fig.:fullsetup}(b)]. In what follows, we highlight three
special positions of the chemical potential, denoted by the black
arrows in the inset of Fig.~\ref{fig.:fullsetup}(b), and label them as
the edge state crossing ($\mu_{\textrm{EC}}$), the Dirac point
crossing ($\mu_{\textrm{DPC}}$), and the bulk crossing
($\mu_{\textrm{BC}}$).

From now on, we assume that the system has  OBC
along both $x$ and $y$ axes. We further assume that the TI is proximity
coupled to a SC, inducing an $s$-wave type of superconducting pairing
of the strength $\Delta$, which, without loss of generality, can be
taken to be positive. We also consider the effect of a magnetic
strip,  placed on one edge of the system [red slab in
Fig.~\ref{fig.:fullsetup}(a)], which we choose to be the $y$-edge at
$x=W$. We further assume that the strip has a finite width $W_Z$ and
is placed symmetrically at a distance $L_{Z}$ from the two
$x$-edges. The resulting induced effective Zeeman field points
along the $x$-axis and, for simplicity, is assumed to be non-zero directly under the magnetic strip.

Finally, we describe the model defined above on a square lattice with
the corresponding tight-binding Hamiltonian, reading
\begin{eqnarray}\label{eqn.:hoppingHamiltonian}
  H=&&\sum_{i =1}^{n_W}\sum_{j=1}^{n_ L}\big[ \bar{c}_{ij}^\dagger
       ( \Delta\eta_1 -\mu_{ij} \eta_3 + \{\Delta_{\textrm{Z}+}^{ij} +
       \Delta_{\textrm{Z}-}^{ij} \tau_3 \} \sigma_1 )\bar{c}_{ij}\big] \nonumber \\
  +&&\sum_{i =1}^{n_W}\sum_{j=1}^{ n_L}\big[ \bar{c}_{ij}^\dagger
      (-4B\tau_3+M\tau_3-4D)\eta_3 \bar{c}_{ij} \big] \\
  +&&\sum_{i =1}^{n_W-1}\sum_{j=1}^{ n_L}\big[ \bar{c}_{ij}^\dagger
      (B\tau_3+iA\tau_1\sigma_3/2 +D) \eta_3\bar{c}_{(i+1)j}+\textrm{H.c.}\big] \nonumber \\
  +&&\sum_{i =1}^{n_L}\sum_{j=1}^{n_L-1} \big[\bar{c}_{ij}^\dagger
      (B\tau_3 + iA\tau_2/2 + D)\eta_3\bar{c}_{i(j+1)}+\textrm{H.c.}\big]\nonumber \textit{,}
\end{eqnarray}
where
$\bar{c}_{ij}^\dagger=(c_{E\uparrow}^\textbf{} ,
c_{H\uparrow}^\textbf{} , c_{E\downarrow}^\textbf{} ,
c_{H\downarrow}^\textbf{} , c_{E\downarrow}^\dagger,
c_{H\downarrow}^\dagger,-c_{E\uparrow}^\dagger, $
$-c_{H\uparrow}^\dagger)_{ij}$. The operator
$(c_{\tau\sigma}^\dagger)_{ij}$ creates an electron in orbital $\tau$
with spin $\sigma$ on a lattice site $(i,j)$. The Pauli matrix
$\eta_i$ acts in particle-hole space and the upper bounds in the
summations over $i$ and $j$ are defined as $n_W = W / a$ and
$n_L = L / a$.

The first line on the right-hand side of Eq.~(\ref{eqn.:hoppingHamiltonian}) describes the
local on-site terms including the SC pairing as well as the spatially
dependent chemical potential $\mu_{ij}$ and the effective Zeeman field
$\Delta_{\textrm{Z}\pm}^{ij}$. The chemical potential is defined as
\begin{eqnarray}
  \mu_{ij}
  = \left\{
  \begin{array}{ll}
    \mu,
    &
      L_{\textrm{Edge}} / a < i < n_W - L_{\textrm{Edge}} / a\ \textrm{and} \\
    &
      L_{\textrm{Edge}} / a < j < n_L - L_{\textrm{Edge}} / a \\
    \mu + \mu_{\textrm{Edge}},
    & \textrm{otherwise}.
  \end{array}
  \right.
\end{eqnarray}
and the Zeeman field as
\begin{eqnarray}
  \Delta^{ij}_{\textrm{Z}\pm}
  =\left\{
  \begin{array}{ll}
    \Delta_{\textrm{Z}\pm}, \hspace{15pt}
    &  i >  n_W - W_{\textrm{Z}} / a\ \textrm{and} \\
    & L_{\textrm{Z}} / a < j < n_L - L_{\textrm{Z}} / a \\
    0,
    & \textrm{otherwise},
  \end{array}
  \right.
\end{eqnarray}
where
$\Delta_{\text{Z}\pm}=(g_\text{E}\pm g_\text{H})\mu_{\textrm{B}} B_x/2$,
$\mu_{\textrm{B}}$ is the Bohr magneton, $g_\text{E,H}$ are the
orbital $g$-factors, and $B_x$ is the effective strength of the magnetic field
generated under the strip.  Previous work~ \cite{wimmer2018} has shown that
$g_\text{E}\gg g_\text{H}$ and $g_\text{H}\approx 0$, thus
$\Delta_{\text{Z}\pm}\equiv \Delta_{\text{Z}}$. For
simplicity, we take $g_\text{E}=1$ for our calculations. The last
three lines of Eq.~(\ref{eqn.:hoppingHamiltonian}) describe the
discretized BHZ Hamiltonian given by Eq. (\ref{eqn:fullBHZ}).

\section{Topological Phase diagram}\label{sec.:sectionIII}

To begin with, we first give a brief summary of the theoretical predictions
for a simplified model in which the counter-propagating helical edge states can be described by an effective one-dimensional low-energy theory. This model works well, if the DP is in the middle of the bulk gap such that there are no hybridization effects between  bulk and edge states. The corresponding Hamiltonian in the basis of fermion operators
$\Phi^\dagger(x)=[\psi^\dagger_\uparrow(x),\psi^\dagger_\downarrow(x),\psi_\downarrow^\textbf{}(x),-\psi_\uparrow^\textbf{}(x)]$
can be written as
\begin{equation}
  \hat{H}=\frac{1}{2}\int dx\
  \Phi^\dagger(x) \mathcal{H} \Phi(x)\textit{,}
\end{equation}
where the Hamiltonian density in momentum space is given by
\begin{eqnarray}\label{eqn.:hamilton-density}
\mathcal{H}=\hbar v_F \hat{k}\sigma_3\eta_3 -\bar{\mu}  \eta_3+\bar{\Delta} \eta_1+\bar{\Delta}_\text{Z}\sigma_1 \textit{.}
\end{eqnarray}\\
Here $\hat{k}=-i\partial_x$ is the momentum operator in real space along the direction of the edge and $v_F$ is the Fermi velocity of the edge modes at the chemical potential $\bar{\mu}$. The fermionic creation operator $\psi^\dagger_\sigma(x)$ acts on an electron with the spin $\sigma$ at the position $x$. 
We also incorporate $s$-wave proximity-induced superconductivity of the strength $\bar{\Delta}$ as well as an effective Zeeman
field of the strength
$\bar{\Delta}_\text{Z}$, which opens a gap at the Dirac point. 
By calculating the energy spectrum for the Hamiltonian in Eq.~(\ref{eqn.:hamilton-density}) and by evaluating the energy gap at $k=0$ ~\cite{fukane2009}, one can identify the topological phase transition that occurs at $\bar{\Delta}_\text{Z}=\bar{\Delta}_\text{Z}^\text{c}$ with
\begin{equation}\label{eqn.:gap}
\bar{\Delta}_\text{Z}^\text{c}=\sqrt{\bar{\Delta}^2+\bar{\mu}^2}.
\end{equation}
The trivial (topological) phase is identified with the regime $\bar{\Delta}_\text{Z}<\bar{\Delta}_\text{Z}^\text{c}$ $(\bar{\Delta}_\text{Z} > \bar{\Delta}_\text{Z}^\text{c})$. The interface between the two topologically distinct regions hosts a MBS \cite{fukane2009}. In connection with our lattice model, the corresponding
interface could emerge at the ends of the magnetic strip [see
Fig.~\ref{fig.:fullsetup}(a)].

\begin{figure}
\centering
\includegraphics[width=1.\linewidth]{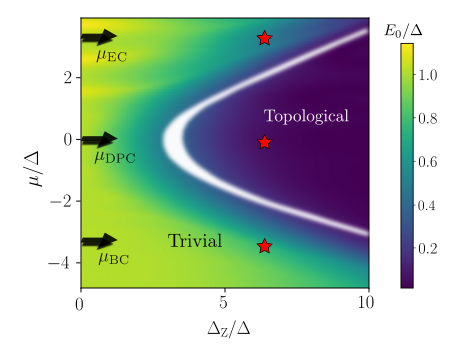}
\caption{Topological phase diagram as a function of $\Delta_\text{Z}$ and the
  chemical potential $\mu$ found numerically for the finite size
  system ($n_W=75$ and $n_L=150$). The color indicates the energy of the
  lowest state $E_0$. In the topological phase (blue area), the
  system hosts MBSs, $E_0\approx 0$. In the trivial phase (green
  area), the lowest state corresponds to a gapped edge or bulk
  state. In particular, for small $\Delta_\text{Z}$,
  $E_0\approx \Delta$. The white blurred line is the phase
  boundary, found numerically from a gap closing condition at
  $k_y=0$ with PBC. The parameters are chosen to be the same as in Fig.~\ref{fig.:fullsetup} with $W_\text{Z}/a=L_\text{Z}/a=10$. The red stars mark three points in the parameter space, on which we elaborate more in detail in Sec.~\ref{sec.:sectionIII}. With this figure, we confirm that the MBSs can emerge also in the system with hidden DPs, in which bulk and edge states are strongly hybridized at the DP. Also in this case, the phase boundary has a parabolic shape [see Eq. (\ref{eqn.:gap})] and its extremum can be used to find the position of the DP. }
\label{fig.:kypd005}
\end{figure}

However, if the DP of a TI is hybridized with bulk states, it is not a priori clear that one can still generate MBSs predicted by the effective model above. At the interface between magnetic and 
superconducting regions, the momentum conservation is lifted. Consequently, an analytical treatment of the problem that would include a hybridization of bulk and edge states is challenging. As well as it is not clear if such a hybridization will not act as efficient source of overlap between MBSs \cite{klino2013}.
Therefore, we tackle these questions numerically within
the framework of a tight-binding model given by
Eq.~(\ref{eqn.:hoppingHamiltonian}). In this model, the SC opens a gap in the
spectrum of size $\Delta$, thus, both bulk and edge states, if present at
the Fermi level, are gapped out. Following the standard reasoning, a local Zeeman field closes the superconducting gap, which, in our case, happens only along one edge of
the sample. Depending on the width of the magnetic sector $W_\text{Z}$, one can ensure that the Zeeman field affects the edge states much more than the bulk states. A wider magnetic strip would suppress the superconductivity in the bulk and the spectrum would become trivially gapless, excluding any possibility to generate bound states.  The separate treatment of edge and bulk states is a desirable feature, which has been investigated in recent experiments~\cite{jaeck2019,bocquillon2020}. 

Diagonalizing numerically Eq.~(\ref{eqn.:hoppingHamiltonian}) while changing both $\mu$ and $\Delta_\text{Z}$, we  generate the entire topological phase diagram, see Fig.~\ref{fig.:kypd005}. 
The corresponding phase diagram can also be compared with the gap closing condition at $k_y = 0$, obtained by solving the problem in a geometry with OBC (PBC) along the $x$ ($y$) directions, which determines a sharper boundary between different phases. In general, we find good agreement between the two results.
In addition, we note that the phase boundary follows the expected parabolic shape,
suggested by Eq.~(\ref{eqn.:gap}).
However, for our set of parameters, the critical Zeeman field, required to reach the topological phase transition, gets renormalized by a factor $\alpha = 3.5$. Such a mismatch with respect to the low energy theory arises due to several reasons. First, the $g$-factor in the ${H}$ band is much smaller than in the ${E}$ band. Second, the magnetic strip covers the edge states only partially, so the effective $g$-factor should be scaled down by the overlap factor (see below). In order to observe the MBSs, it is most favourable to place the
chemical potential $\mu$ exactly at the DP, since in this case,
$\Delta_\text{Z}$ required to enter the topological phase is 
smallest, see Fig.~\ref{fig.:kypd005}. This trick consequently allows one
to deduce the exact position of the DP by exploring the phase diagram and to determine whether it is
hidden or not~\cite{molenkamp2015, sullivan2015, songbo2014, wimmer2018}.

\begin{figure}[t]
\centering
\includegraphics[width=1\linewidth]{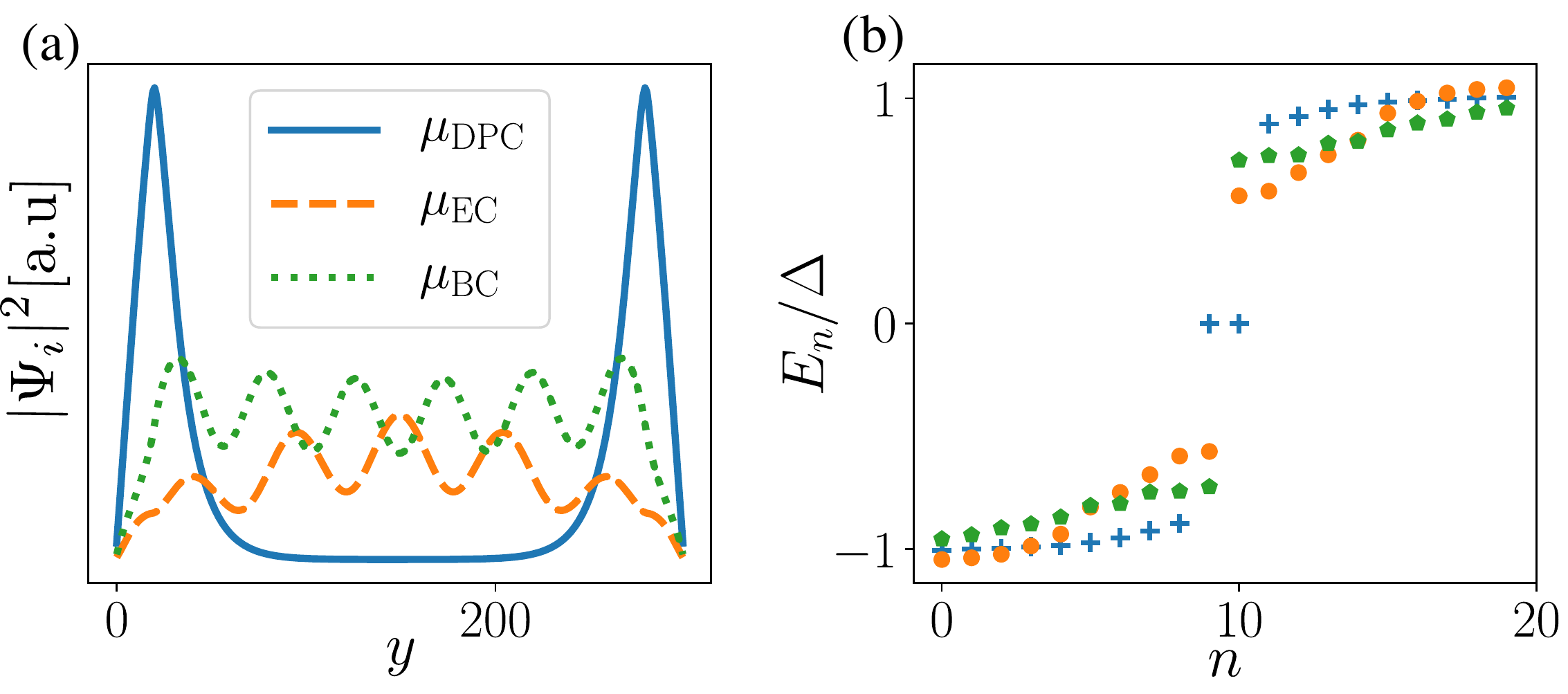}
\caption{(a) Probability density of the lowest energy state
  evaluated along the $y$ edge at $x=W$ for three different positions of
  the chemical potential $\mu$. Only for
  the chemical potential   tuned  into the topological phase, i.e., $\mu=\mu_{\rm DPC}$,  the lowest energy states correspond to zero-energy MBSs (blue solid line), localized at the two ends of the magnetic
  strip.  Outside of the topological phase, the lowest energy state
 is at the finite energy $E_0 > 0$. For
  $\mu_{\rm EC}$  ($\mu_{\rm BC}$) shown by the orange dashed line (dotted green lines), the state is localized under the
   magnetic strip (spread over the entire system). (b) The 20 lowest energy states calculated for the
  values of the chemical potential specified in the panel (a). A pair
  of zero-energy states emerge for $\mu = \mu_{\rm DPC}$. The
  parameters are the same as in Fig.~\ref{fig.:kypd005} and the
  value of the Zeeman field is fixed to $\Delta_\text{Z}/\Delta=7.5$.}
\label{fig.:MBS}
\end{figure}

To emphasize our statement, we also investigate the low-energy spectrum and the probability densities
of the lowest energy state for three positions of the chemical
potential, see Fig.~\ref{fig.:MBS}. Indeed, in the topological phase, the MBSs are localized at the interface between  magnetic and superconducting regions and protected from the hybridization by the corresponding gaps in the spectrum of bulk and edge states. The two-dimensional plots of the MBS probability density are given in the Appendix \ref{Sec.:App2}.
In the trivial phase, the lowest energy state is either
spread over the entire sample or is localized under the magnetic strip. 

\section{Spin and Charge polarization}\label{sec.:sectionIV}
Besides the direct measurement of MBSs, an independent signature of the topological phase transition can be obtained by focusing on the spin and charge properties of our system~\cite{pawel1,serina2018,Thakurathi2020}. Such studies can help to exclude a possibility of identifying a trivial phase as a topological due to the observation of a stable zero-energy bias peak arising from the trivial Andreev bound state \cite{a1,a2,a3,a4,a5,a6,a7,a8}.
Needed techniques have already been successfully employed in recent experimental measurements~\cite{lanzara2016,jaeck2019}.
In the following, we focus on the expectation values of the charge and spin operators defined as $Q = \eta_3$ and $\mathbf{S}=(S_x, \eta_3 S_y, S_z)$ with $S_\mu=[1/2(1+\tau_z)+3/2(1-\tau_z)]\sigma_\mu/2$,
respectively.  
Here, the factor $1/2$ ($3/2$) correspond to the spin of the $E$ ($H$) band  in the BHZ model. Since our model is defined on a lattice, the expectation
values of these operators depend explicitly on the lattice position ($i,j$)
and are defined as
\begin{eqnarray}\label{eq.:localspinandcharge}
  &&\braket{\mathbf{S}(i,j)}_{n}=
     \bra{\psi_n(i,j)} \mathbf{S} \ket{\psi_n(i,j)},
  \notag \\
  &&\braket{Q(i,j)}_{n}=
     \bra{\psi_n(i,j)} \eta_3 \ket{\psi_n(i,j)},
\end{eqnarray}
where $\psi_n(i,j)$ is an eigenstate with eigenenergy $E_n$, obtained by diagonalizing the lattice
model of Eq.~\eqref{eqn.:hoppingHamiltonian}. The spin (charge) operator is measured in units of $\hbar/2$ (the electron charge
$|e|$). Moreover, to get a clear distinction between bulk and edge states, we define the average spin (charge) polarization  of the
system by integrating over a domain $\mathcal{D}$ defined as an area under and around  the
magnetic strip:
\begin{align}\label{eq.:spinandcharge}
  \braket{\mathbf{S}}_n
  &=
    \frac{1}{A_{\mathcal{D}}}
    \sum_{(i,j) \in \mathcal{D}} \braket{\mathbf{S}(i,j)}_{n},
    \notag \\
  \braket{Q}_n
  &=
    \frac{1}{A_{\mathcal{D}}}
    \sum_{(i,j) \in \mathcal{D}} \braket{Q(i,j)}_{n}. 
\end{align}
Here, the sum runs over all the sites inside the domain $\mathcal{D}$
and the total sum is divided by the area of this domain equal to $A_{\mathcal{D}}$.

\begin{figure}
\centering
\includegraphics[width=0.8\linewidth]{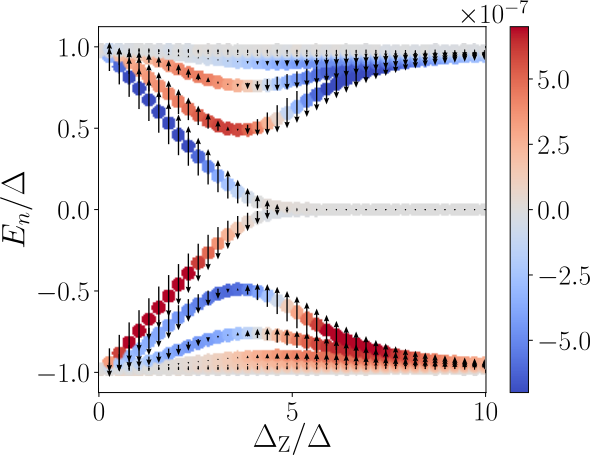}
\caption{The 20 lowest eigenenergies $E_n/\Delta$ as a function of Zeeman field $\Delta_Z/\Delta$ (normalized by the gap $\Delta$) obtained numerically at  $\mu = \mu_\text{DPC}$.  The color code denotes the charge polarization $\braket{Q}_n$ (see Eq.~\ref{eq.:spinandcharge}). The length and direction of the black arrows correspond to the spin polarization along the $x$ axis $  \braket{S_x}_n$. For the low-energy edge states, both observables flip their
 sign across the topological phase transition. As expected, the bulk states remain unaffected by the local magnetic field. The parameters are the same as in Fig.~\ref{fig.:kypd005} and the domain $\mathcal{D}$ is chosen to be twice as wide as the magnetic strip. }
\label{fig.:spinx}
\end{figure}

In Fig.~\ref{fig.:spinx} we show the numerical result of the calculation of
the lowest eigenvalues $E_n$
as a function of the ratio $\Delta_{\mathrm{Z}} / \Delta$ as the
system moves across the topological phase transition. There, we present the spin-$x$ component $\braket{S_x}_n$ (black arrows) of the averaged spin polarization
$\braket{\mathbf{S}}_n$ as well as the charge $\braket{Q}_n$, encoded in the color of the
points. The two other spin polarization components are vanishingly small. We distinguish two different types of contributions: the
first one coming from the edge states lying inside the superconducting
gap $|E_n|<\Delta$ and the second one coming from the bulk states at
energies $|E_n| \ge \Delta$ exceeding the superconducting gap. The spin polarization of the edge states
flips its sign across the topological phase transition, whereas the one of the bulk states remains
unaffected by the effective Zeeman field. Moreover, the signal under the magnetic strip is mainly determined by edge states rather than the bulk states. The two 
zero-energy states do not carry any polarization or charge, which is in agreement with the properties of MBSs \cite{composite}.

\begin{figure}[t]
\centering
\includegraphics[width=1.0\linewidth]{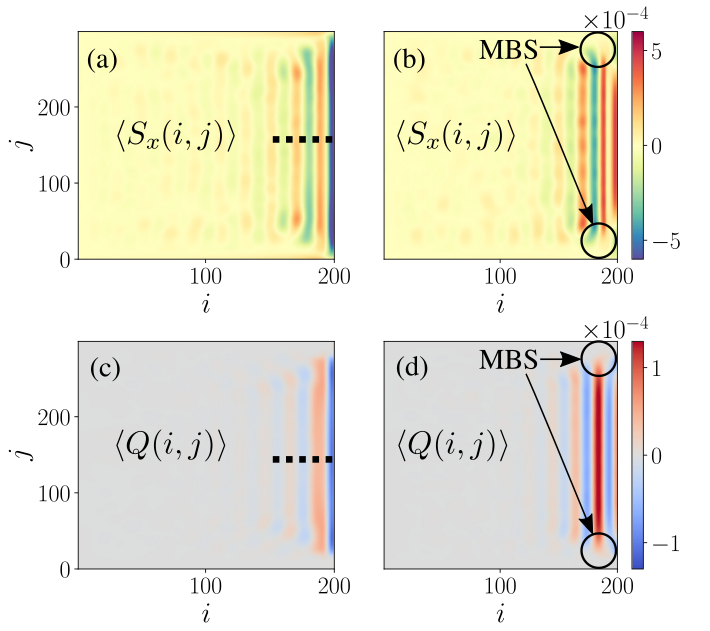}
\caption{Spatially resolved spin (charge) polarization
  $\braket{S_x(i,j)}$ $[\braket{Q(i,j)}]$ determined by taking contributions of five lowest energy states below the chemical potential. The chemical potential is tuned to $\mu_\text{DPC}$. The panels (a) and (c) [(b) and (d)] correspond to the trivial (topological) phase at $\Delta_\text{Z}/\Delta=3.5$ [$\Delta_\text{Z}/\Delta=7.5$]. We find that during the topological phase transition
  both quantities undergo a sign flip close to the edge, where the dominant contribution to the total polarization is localized. In the topological phase,   $\braket{S_x(i,j)}$ and $[\braket{Q(i,j)}]$ are strongly suppressed at the ends of the magnetic strip where MBSs are localized. The cross-section plots, denoted by the black dashed lines, are presented in Appendix~\ref{Sec.:App3}. The remaining set of parameters is the same
  as in Fig.~\ref{fig.:kypd005}.}
\label{fig.:spinedge}
\end{figure}

To underline these results, in Fig.~\ref{fig.:spinedge}, we also study 
spatially resolved spin and charge polarization, which are defined as
\begin{eqnarray}\label{eq.:localspinXandcharge}
    \braket{S_x(i,j)}=\sum_{E_n<0}\braket{S_x(i,j)}_{n} ,\nonumber \\
 \braket{Q(i,j)}=\sum_{E_n<0}\braket{Q(i,j)}_{n},
\end{eqnarray}
where the sum runs over some finite interval of negative energies below the chemical potential with the goal in mind to mimic in this way the energy resolution of STM tips. We note here that the number of states over which we perform the summation does not play a crucial role,
since  the contribution of the bulk states delocalized over the entire sample is vanishingly small, see Fig. \ref{fig.:spinx}.
By taking a few of the energetically lowest lying eigenstates, we calculate the spin polarization in the trivial phase [see Fig.~\ref{fig.:spinedge}(a)], which corresponds in total to negative values of the polarization $\braket{S_x}_n$ [see Fig.~\ref{fig.:spinx}(a)]. 
The main contribution to the spin polarization is mainly localized at the edge of the system with decaying oscillations into the bulk. A similar pattern is observed in the topological phase as shown in Fig.~\ref{fig.:spinedge}(b). Most interestingly, the sign of the dominant contribution of the spin polarization close to the edge flips (see App.~\ref{Sec.:App3} for more details). Moreover, its intensity is strongly suppressed in the area where two MBSs are localized. This effect arises due to the vanishing polarization of MBSs mentioned above.

Similarly, we present in Fig.~\ref{fig.:spinedge}(c,d) the results of the calculation of the spatially resolved charge polarization in the trivial and topological phases. We observe that it behaves in the same way as the spin polarization with the dominant contribution at the edge flipping its sign across the topological phase transition point. Furthermore,  we again observe  a suppression of the charge polarization in the area of the localization of MBSs.

The observables defined above can be accessed with scanning tunneling microscope (STM) techniques~\cite{lanzara2016,jaeck2019}. Conclusively, an STM with a non-magnetic tip would allow one to probe the charge expectation value $\braket{Q(i,j)}$, while a spin-polarized STM would give access to
$\braket{S_x(i,j)}$ \cite{pawel,ali}. Thus, these techniques open up a path to  observe the  topological phase transition accompanied by a detection of the position of the DP in the spectrum.

\section{Disorder effects and increase in the magnetic stripe width}\label{sec.:sectionV}

In this section, we demonstrate the stability of the predicted effects towards disorder.  More precisely, we study the properties of the system perturbed by a local on-site variations of the chemical potential. 
%We neglect faults in the superconducting gap $\Delta$, since it is assumed to have weak and smooth variation. 
The fluctuations in the chemical potential are assumed to have the form $\mu +\delta\mu$, where $\delta\mu$ is sampled from a Gaussian distribution with zero mean and a standard deviation $\sigma_\mu$. 

\begin{figure}
\centering
\includegraphics[width=1.0\linewidth]{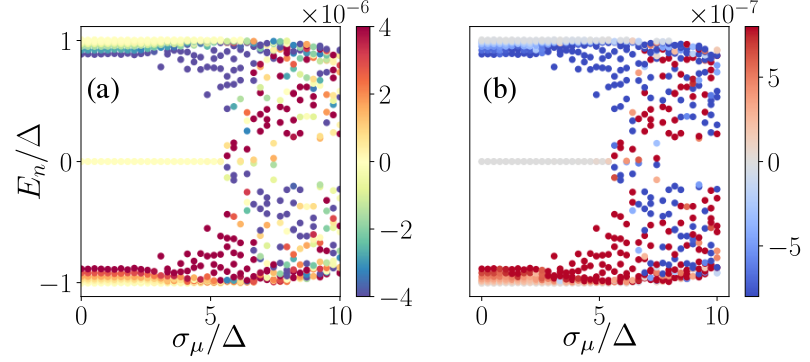}
\caption{The spectrum of the 20 lowest-lying energy states in the topological phase ($\Delta_\text{Z}/\Delta = 7.5$) as a function of the disorder strength in the chemical potential $\sigma_\mu$. The color code
  denotes (a) the spin polarization along the $x$ axis  and  (b) the charge polarization. At disorder strength of several $\Delta$, the bulk gap closes. Below this value the polarization of bulk states is well-defined and can be used to distinguish between topological and trivial phases. All non-zero
  parameters are the same as in Fig.~\ref{fig.:kypd005}.}
\label{fig.:dis_PD}
\end{figure}

\begin{figure}
\centering
\includegraphics[width=1.0\linewidth]{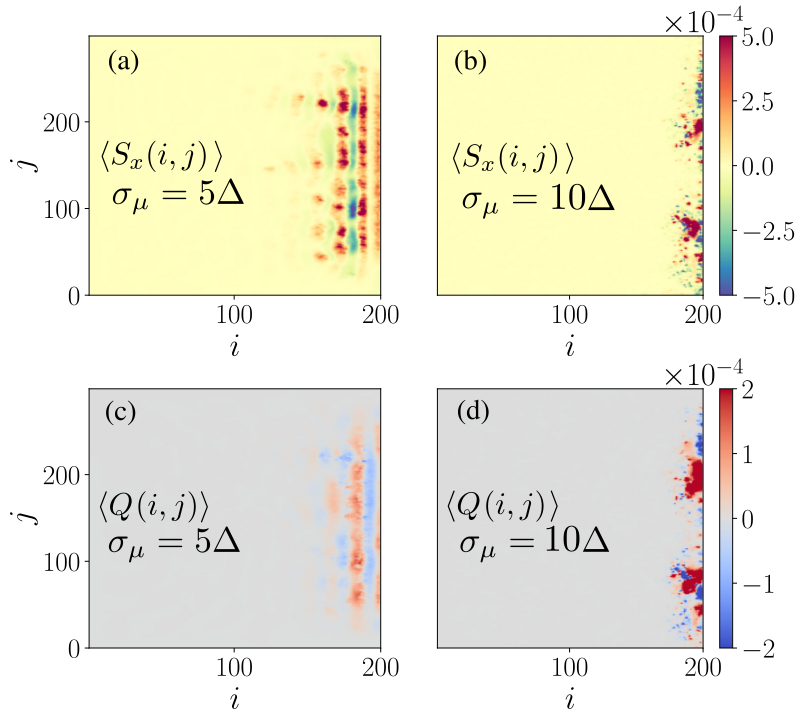}
\caption{Spatially resolved spin (charge) polarization $\braket{S_x(i,j)}$ $[\braket{Q(i,j)}]$ in the presence of disorder in the chemical potential is shown in panels (a,b) [(c,d)].   (a,c) In the weak disorder regime, the bulk gap is still open and the polarization patterns are similar to the ones in the clean case.  (b,d) Such features disappear in the strong disorder regime where the bulk gap is closed. The set of parameters is the same
  as in Fig.~\ref{fig.:dis_PD}.}
\label{fig.:dis_PD2D}
\end{figure}

First, we fix  the value of the effective Zeeman field such that the system is in the topological phase.
Then, we increase the disorder strength and, similarly to Fig.~\ref{fig.:spinx}, we calculate the lowest eigenvalues $E_n$, as well as the spin and charge expectation values $\braket{\mathbf{S}_x}_n$ and $\braket{Q}_n$, respectively, of the corresponding eigenstates as a function of the ratio $\sigma_\mu / \Delta$, see Fig.~\ref{fig.:dis_PD}.  We find that the MBSs remain stable and well separated from the bulk states for  disorder strengths  up  to about $\sigma_\mu=10\Delta$. For stronger disorder, the bulk gap closes and the featues in spin and charge polarization can no longer be distinguished. In general, the larger is the topological gap, the more robust is the system under this kind of disorder. Similarly, we can study effects of disorder on the spatial profile of the spin and charge polarization, see  Fig.~\ref{fig.:dis_PD2D}. In the topological phase and for weak disorder, the spin and charge polarization oscillate and decay into the bulk [see Fig.~\ref{fig.:dis_PD2D}(a,c)] as we already observed in the clean case. These features disappear in the limit of strong disorder as the bulk gap closes, see Fig.~\ref{fig.:dis_PD2D}(b,d).

\begin{figure}[t]
\includegraphics[width=0.9\linewidth]{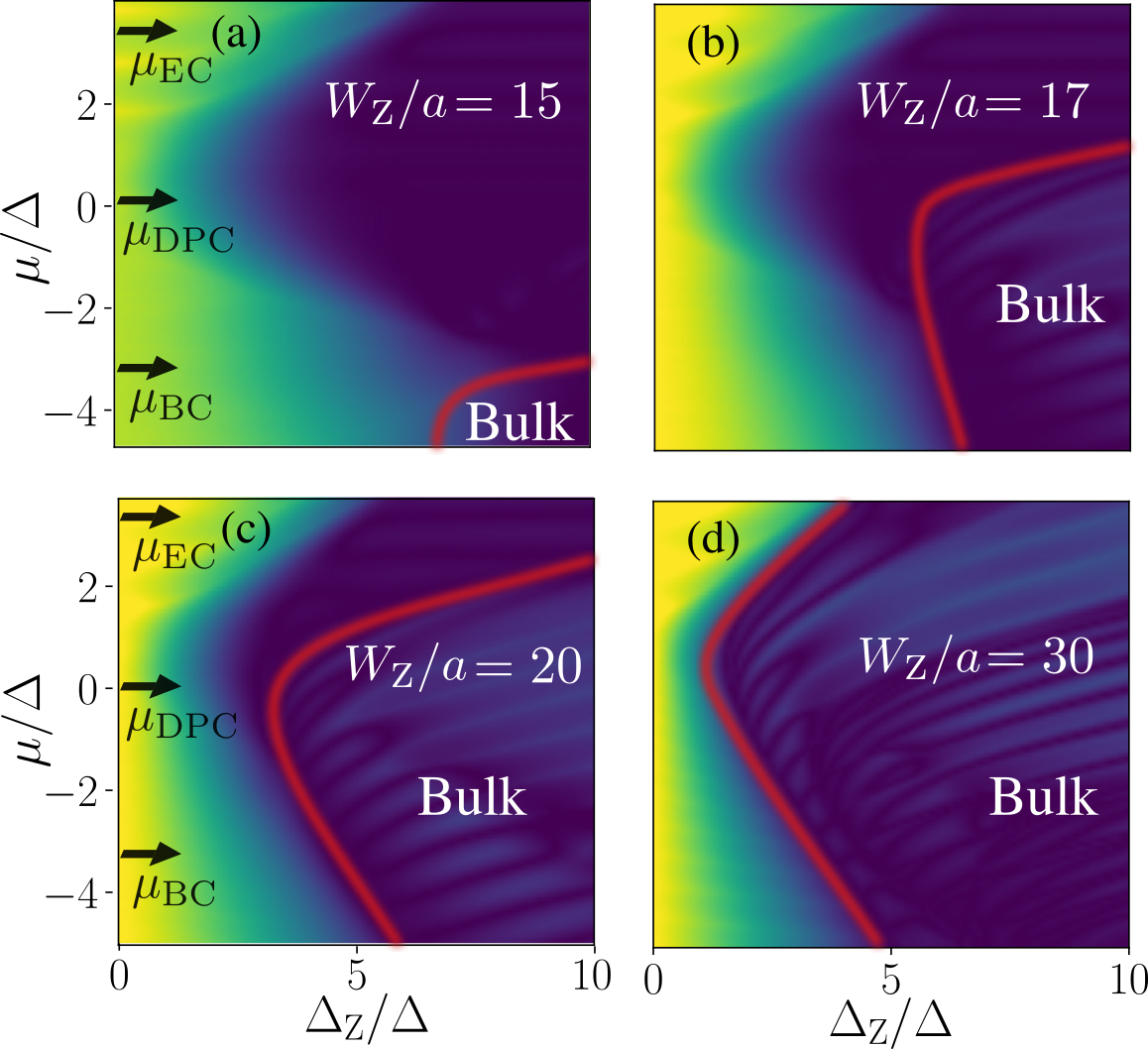}
\caption{Topological phase diagram as a function of $\Delta_\text{Z}$ and the
  chemical potential $\mu$ obtained for different widths of the magnetic strip: (a) $W_\text{Z}/a=15$, (b) $W_\text{Z}/a=17$, (c) $W_\text{Z}/a=20$, and (d) $W_\text{Z}/a=30$. We find that the unique features associated with the topological phase at the system edge get smeared out as a result of the increased coupling between the edge and the bulk states as well as a result of the suppression of the superconductivity induced in the bulk states at the edge. The wider the magnetic strip,  the stronger these effects are. We use the red lines to outline the region of the phase diagram, where the bulk effects dominate and the edge features of the topological phase get lost. The rest of the parameters are the same as in Fig.~\ref{fig.:kypd005} as well as the color code.}
\label{fig.:PD-strip s}
\end{figure}

Finally, we comment on how the width of the magnetic strip $W_\text{Z}$ affects the previously presented results. As noted above, the small values of $W_\text{Z}$ (much smaller than the localization length of edge states) is not favourable since in this case the effective $g$-factor would be strongly suppressed as the effective magnetic field would be acting only on the very small part of the edge state. However, very wide strips are also not favourable.
Indeed, one can expect that by increasing $W_\text{Z}$ one also increases mixing between bulk and edge states, which must destroy the topological phase. In order to address this question more quantitatively, we perform numerical simulations for different $W_\text{Z}/ a$   varying from 15 to 30 sites. For each value of $W_\text{Z}$ we calculate the phase diagram similarly to Fig.~\ref{fig.:kypd005}. The resulting phase diagrams are summarized in Fig.~\ref{fig.:PD-strip s}. First, we notice that by increasing $W_\text{Z}$ the minimal value of the effective Zeeman field $\bar{\Delta}_\text{Z}^\text{c}$ for which the topological phase is achieved is indeed decreasing. For wide strips, $\bar{\Delta}_\text{Z}^\text{c}$ comes close to $\Delta$ for the chemical potential tuned to the DP. The parabolic shape is also preserved at small fields but the curvature is increased due to larger effective $g$-factor caused by the fact that the larger area of edge state is covered by  the magnetic strip.  However, we also find that an increase of $W_\text{Z}/a$ blurs the previously unique signatures into a partly non-distinguishable superposition of bulk and edge states such that the system is gapless at the edge. The wider the strip is, the more effect it has on the bulk states. As the topological phase is achieved for $\Delta_Z$ greatly exceeding the superconducting pairing $\Delta$, the supercounductivity in the bulk states gets locally suppressed under the magnetic strip,  giving rise to delocalized bulk modes at zero energy. These gapless bulk states mix with gapped edge states. First, this leads to the overlap between MBSs, and subsequently, destroys the topological phase completely.
Thus, it is preferable to keep the effective  Zeeman field to be confined on the length scale of the localization length of edge states, $\chi$ (see
Appendix~\ref{Sec.:App1}). 

\section{Conclusion}
To conclude, in this work we considered a modified BHZ model to 
investigate TIs with hidden Dirac point or with non-uniform chemical potential at the edge of the TI, which allows one to address bulk and edge states separately. The system under consideration  is proximity-coupled to an s-wave SC and to a magnetic strip of varying width that generates an effective Zeeman field on the edge of the TI. We showed that robust MBSs can be generated in the setup independent of the location of the DP in the spectrum. The topological phase diagram associated with the MBSs allows one to determine precisely the position of the DP, even when it is hidden deep inside the energy bulk states. We propose that local measurement techniques, such as STM or polarized-light measurements, can be used to detect the topological phase transition independent of the appearance of the zero-bias peak originating from MBSs. The local spin and charge polarizations flip their sign at the topological phase transition point. Even if such signatures are most pronounced in materials with non-hidden DPs, where the bulk and edge states are well separated, the proposed effects are still present even if the DP is hidden. Remarkably, we find that these features are very stable against weak disorder.

We thank T.~H.~Galambos, F.~Ronetti, and Y.~Volpez for fruitful discussions. This work was supported by the Swiss National Science Foundation, NCCR QSIT, and the Georg  H.~Endress foundation. This project received funding from the European Unions Horizon 2020 research and innovation program (ERC Starting Grant, grant agreement No 757725). 

\appendix

\section{Hybridization between edge  and bulk states}\label{Sec.:App1}
\begin{figure}[H]
\centering
\includegraphics[width=0.8\linewidth]{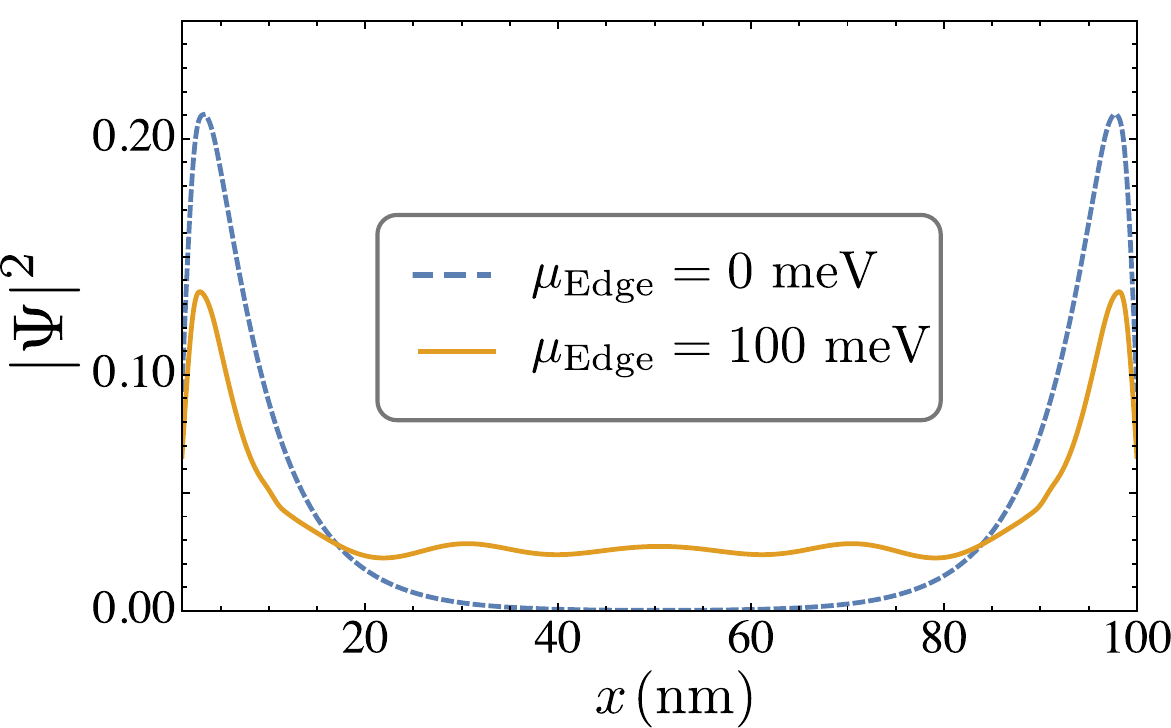}
\caption{
The probability density of the four lowest energy states at the DP. For $\mu_\text{Edge}=0$ meV (dashed blue), the DP is inside the TI bulk gap. The corresponding wavefunctions are clearly localized on the right and left edges and are clearly separated from bulk states. For $\mu_\text{Edge}=100$ meV (solid orange), the DP is hidden. The wavefunctions have contributions from both bulk and edge states, which demonstrates  strong  hybridization between them. The parameters are the same as in Fig.~\ref{fig.:fullsetup}(b).
}
\label{fig.:Appendix-decay}
\end{figure}

In this Appendix, we highlight the hybridization between edge and bulk states that occurs when the DP is hidden in the bulk ($\mu_\text{Edge}=100$ meV) and compare obtained results with the case of a non-hidden DP ($\mu_\text{Edge}=0$ meV).
We consider the model of an isolated TI (without SC or magnetic strip being attached) in a geometry with OBC along the $x$-direction and PBC along the $y$-direction. The energy dispersion for $\mu_\text{Edge}=100$ meV is shown, for example, in Fig.~\ref{fig.:fullsetup}(b): the DP is hidden. In contrast to that, for $\mu_\text{Edge}=0$ meV, the DP is in the middle of the bulk gap. In this geometry, at $k_y=0$, the spectrum has a four fold degeneracy where we also account for the degeneracy between two edges. To estimate the localization length of the edge states, we plot in Fig.~\ref{fig.:Appendix-decay} the probability density of the four lowest energy states at the DP, for both $\mu_\text{Edge}=0$ meV and $\mu_\text{Edge}=100$ meV. 

From Fig.~\ref{fig.:Appendix-decay}, we conclude that the decay length of the edge states is of the order of 10 sites for this choice of parameters. This is in agreement with the analytical estimations of the decay length, which is equal to $\chi/a=-A/M\approx 9$. Hence, in the calculations in the main part we took the width of the magnetic strip to be equal to $W_\text{Z}/a=10$ sites. When the DP is hidden and overlaps in energy with  bulk states, one clearly sees the hybridization between edge and bulk states. The hybridization strength increases with $\chi$, since the more an edge state leaks into the bulk, the larger the overlap between edge and bulk states.

\section{Probability density of the MBSs}\label{Sec.:App2}
\begin{figure}[t]
\centering
\includegraphics[width=0.9\linewidth]{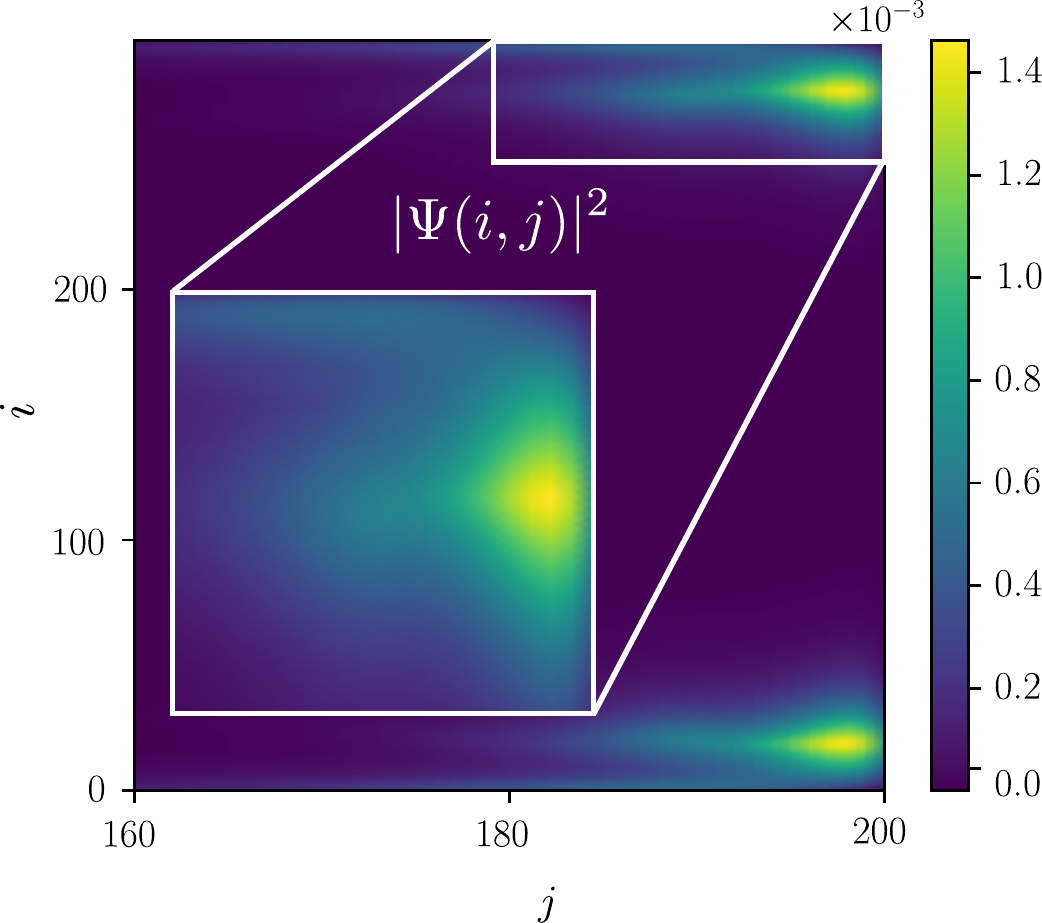}
\caption{The probability density for the two zero-energy modes from Fig.~\ref{fig.:MBS}(b). The inset shows a zoom into the MBS at the upper right edge, close to the ends of the magnetic strip. The MBSs are localized at the end of the magnetic strip and extend the most along the edge in $x$-direction. The non-zero parameters are the same as in Fig.~\ref{fig.:MBS}.}
\label{fig.:MBS3D}
\end{figure}

This section shows the result of the calculation of the MBSs wavefunction (the same as in Fig.~\ref{fig.:MBS}) which is expressed in the full system, instead of the edge $x = W$ only. Since the edge state of an unperturbed TI has a finite width, we expect that the MBS decays in the two topological distinct regions, as well as along the finite interface, which is shown in Fig.~\ref{fig.:MBS3D}.  There, we find as expected the two MBSs on both ends of the magnetic strip. This reflects the one-dimensional character of the TI edge state, since the MBS probability density peaks  at the same positions where the edge state probability density is maximum close to the interface.

\begin{figure}[b]
\centering
\includegraphics[width=0.9\linewidth]{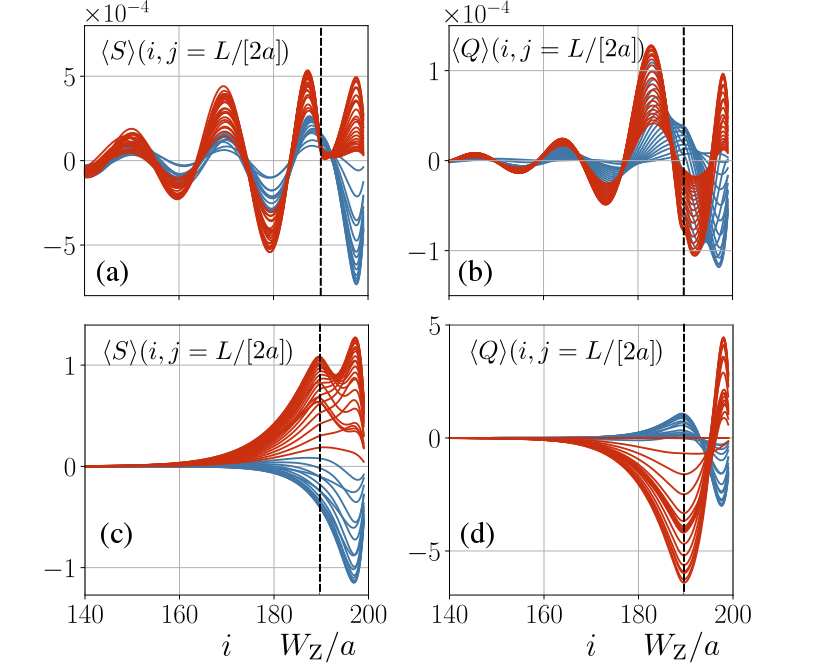}
\caption{Cross-section of the spin (left panels) and charge (right panels) polarizations  for (a,b) a hidden DP and (c,d) a non-hidden DP. The Zeeman field varies from $0.25\Delta$ to $10\Delta_\text{Z}$ across the topological phase transition.
The trivial phase is shown in blue, while the topological phase is shown in red. The sign-flip of the polarization is  most pronounced in the spin polarization and observed most strongly directly under the magnetic strip and in the case of a non-hidden DP. The hybridization between bulk and edge states suppresses the spin-flip away from the strip. The black dashed lines mark the edge of the magnetic strip. The rest of the parameters is the same as in Fig.~\ref{fig.:spinedge}.}
\label{fig.:cross}
\end{figure}

\section{Cross section of spin and charge expectations}\label{Sec.:App3}
In this appendix, we present a more detailed study of the spatial dependence of the spin and charge expectation values  $\braket{S_x (i,j)}$ and $\braket{Q(i,j)}$, defined in Eq.~(\ref{eq.:localspinXandcharge}) of the main text. We focus on a particular cross-section of our system at $y = L / 2$ (see dotted line in Fig.~\ref{fig.:spinedge}). Moreover, we compare the two different situations, corresponding to the case when the DP of the TI is in the middle of  the bulk TI gap (which is obtained by taking $\mu_{\textrm{Edge}} = 0$) and when it is hidden at the energies of the bulk states (with $\mu_{\textrm{Edge}} = 100$ meV). For both scenarios, we assume that the chemical potential is tuned to the DP. The result of such a calculation is shown in Fig.~\ref{fig.:cross}. There, we vary for every situation the value of the effective Zeeman field and superimpose the resulting plots. The results obtained in the topologically trivial phase are shown in blue, while the results in the topological phase are shown in red. We find that both the spin and charge polarization decay into the bulk of the TI as expected. In the case of a hidden DP shown in Fig.~\ref{fig.:cross}(a,b), we observe much slower decay. In addition, as a result of the hybridization between the bulk and edge states, both quantities oscillate and flip their signs only under the strip, whereas, away from the strip, where the bulk states dominate, no sign flip is observed.  We find that outside of the magnetic region, the oscillations of both the spin polarization and charge are in phase for all the values of the Zeeman field. Nevertheless, inside the magnetic region, a sign flip appears in the oscillations of these two quantities in the topological and trivial phases. We notice that this feature allows one to determine very precisely the exact location of the phase transition. In the case of the non-hidden DP,  where only edge states contribute, all signatures of the sign flip are well-pronounced even away from the strip.

\newpage
\bibliography{Literatur}

\begin{thebibliography}{1}
\bibitem{footnote} We note that such a non-uniform shift of the chemical potential can arise due to strong coupling to a thin superconducting layer as was recently shown in one- and two- dimensional Rashba systems \cite{prox1,prox2,prox3,prox4,prox5,prox6}.
\end{thebibliography}

\end{document}